\documentclass[journal=jacsat,manuscript=letter]{achemso}

\usepackage{chemformula} 
\usepackage[T1]{fontenc} 

\usepackage{graphicx}
\usepackage{dcolumn}
\usepackage{bm}
\usepackage{appendix}
\usepackage[utf8]{inputenc}
\usepackage{siunitx}
\usepackage{hyperref}
\usepackage{array}
\usepackage{float}
\hypersetup{
    colorlinks=true,
    linkcolor=black, 
    filecolor=black, 
    urlcolor=black, 
    citecolor = black,
    pdftitle={Overleaf Example},
    pdfpagemode=FullScreen,
    }
    
\usepackage{ulem}
\newcommand{\revision}[1]{\textcolor{black}{#1}} 
\newcommand{\revis}[1]{\textcolor{black}{#1}} 

\title{Interfacial resistive switching by  multiphase polarization in ion-intercalation nanofilms}

\author{Huanhuan Tian}
\affiliation{Department of Chemical Engineering, Massachusetts Institute of Technology, Cambridge, MA 02139, USA}

\author{Martin Z. Bazant}
\affiliation{Department of Chemical Engineering, Massachusetts Institute of Technology, Cambridge, MA 02139, USA}
\alsoaffiliation{Department of Mathematics, Massachusetts Institute of Technology, Cambridge, MA 02139, USA}
\email{bazant@mit.edu}

\begin{document}


\begin{abstract}

Nonvolatile resistive-switching (RS) memories promise to revolutionize hardware architectures with in-memory computing. Recently, ion-interclation materials have attracted increasing attention as potential RS materials for their ion-modulated electronic conductivity.   In this Letter, we propose RS by multiphase polarization (MP) of ion-intercalated thin films between ion-blocking electrodes, in which interfacial phase separation triggered by an applied voltage switches the electron-transfer resistance.  We develop an electrochemical phase-field model for simulations of coupled ion-electron transport and ion-modulated electron-transfer rates and use it to analyze the MP switching current and time, resistance ratio, and current-voltage response. The model is able to reproduce the complex cyclic voltammograms of lithium titanate (LTO) memristors, which cannot be explained by existing models based on bulk dielectric breakdown. The theory predicts the achievable switching speeds for multiphase ion-intercalation materials and could be used to guide the design of  high-performance MP-based RS memories. 

\noindent \textbf{keywords}: resistive switching, phase-field modeling, electron transfer, ion intercalation

\end{abstract}

\maketitle

In the era of Big Data, the transfer of data between the processing unit and memory  has increasingly limited the performance of traditional computing architectures. The "von Neumann bottleneck" can potentially be addressed by in-memory computing, which requires resistive-switching (RS) devices with multiple, nonvolatile resistance states that can be tuned by applied voltages
\cite{ Waser2009Redox-basedChallenges, Jeong2012EmergingStatus,  Ielmini2018In-memoryDevices, Waser2019IntroductionApplications, Li2020, Zidan2018TheSystems}. RS  devices include two-terminal memristors and three-terminal synaptic transistors \cite{Li2020}.  
The switching can be bipolar or unipolar, depending on whether or not the set and reset voltages require different signs, respectively.\cite{Waser2019IntroductionApplications} 
Studied RS mechanisms include ion migration \cite{Sawa2008ResistiveOxides, Waser2009Redox-basedChallenges, Valov2011ElectrochemicalProspects, Ielmini2016ResistiveScaling, Wang2020ResistiveProcessing, Ielmini2018In-memoryDevices, DelValle2018ChallengesComputing},  amorphous–crystalline transition  \cite{ Wang2020ResistiveProcessing, Ding2019Phase-changeOperation, Wong2010PhaseMemory,Ielmini2018In-memoryDevices, DelValle2018ChallengesComputing}, ferroelectricity \cite{Wang2020ResistiveProcessing, Ielmini2018In-memoryDevices}, and tunneling magneto-resistance \cite{Wang2020ResistiveProcessing, Ielmini2018In-memoryDevices}. This work describes an RS mechanism based on interfacial phase separation that is limited by bulk ion migration.

Typically, the ion migration mechanism incorporates an ion-conducting nanofilm, and an active electrode that can inject (consume) active ions or vacancies into (from) the nanofilm. Specifically, the mechanism is called electrochemical metallization (ECM) if metal cations (e.g., Cu$^{2+}$, Ag$^+$) are active and valence change mechanism (VCM) if oxygen vacancies are active \cite{Waser2009Redox-basedChallenges, Li2020}. The ion migration  mechanism can be further divided into the bulk and interfacial types, according to which resistance dominates. The bulk mechanisms often involve the formation of conductive filaments (e.g., Cu, Ag, or oxygen vacancy rich regions) by implantation of ions or vacancies from an active electrode, and the dissolution of the filaments by a reverse process driven by a reverse voltage (bipolar switching) or by Joule heating generated by a larger voltage (unipolar switching) \cite{ Wang2020ResistiveProcessing, DelValle2018ChallengesComputing, Ielmini2016ResistiveScaling, Valov2011ElectrochemicalProspects, Zhang2020High-throughputMemory}.  In the interfacial mechanisms, usually the two electrodes have, respectively, Ohmic contact and Schottky contact with the nanofilm, and the latter is sensitive to the local concentration of ions or vacancies which can be enriched or depleted by electric field (bipolar switching) \cite{Sassine2016Devices, Sawa2008ResistiveOxides, DelValle2018ChallengesComputing, Hur2010ModelingOxides, Yang2008MemristiveNanodevices, Solanki2020InterfacialMemories}. 

In recent years, ion-intercalation materials, which have been widely used for batteries \cite{Nitta2015Li-ionFuture, Stark2019Intercalation2D, Zhou2021LayeredMaterials}, have received increasing attention as novel ion-migration-based RS materials. These materials allow for reversible insertion of ions into the lattice without destroying the original crystal structure \cite{Riess1997ElectrochemistryConductors}, often following a multi-phase mechanism \cite{Bazant2013TheoryThermodynamics}. Their electronic conductivity usually depends on ion concentration, since the inserted/deserted ions usually contribute (nearly) free electrons/holes to the conduction/valence bands \cite{Morgan2016VariationSurfaces, Liu2017TheLi7Ti5O12, Yao2020ProtonicNetworks, Gellings1997TheElectrochemistry, Cox2010TransitionProperties}. This property has been directly used to design bulk-type memristors \cite{Nguyen2018DirectNanobatteries} and synaptic transistors \cite{Sharbati2018Low-PowerComputing, Fuller2017Li-IonComputing, Yao2020ProtonicNetworks, Onen2021CMOS-CompatibleLearning} whose ion concentration is adjusted by ion insertion/desertion through the active or gate electrodes. Compared to ECM or VCM devices, such devices should have good reproducibility and controllability since  the conductivity can be precisely and reversibly controlled by current pulses \cite{Yao2020ProtonicNetworks}.

This work focuses on ion-intercalation memristors enclosed by ion-blocking electrodes, inspired by the LTO (lithium titanate, Li$_{4+3\xi}$Ti$_{5}$O$_{12}$) memristors developed in Ref. \citenum{Gonzalez-Rosillo2020Lithium-BatterySeparation}. LTO is a commonly used anode material for Li-ion batteries,  and follows a two-phase mechanism \revision{and insulator-metal transition }during lithiation ($\xi: 0 \rightarrow 1$). \cite{Zhao2015APerspectives}. The memristors were made of LTO4 ($\xi \approx 0$) or LTO7 nanofilms ($\xi \approx 1$) sandwiched by Pt electrodes, and showed bipolar switching behaviors\revision{. The ion-blocking electrodes make the RS mechanism different from those reviewed before.} Refs.\citenum{Gonzalez-Rosillo2020Lithium-BatterySeparation, Fraggedakis2020} explain the RS of LTO memristors by the formation of conductive filaments by dielectric breakdown, based on a phase-field model including the electrostatic self-energy that depends on the magnitude of the applied potential. \revis{However, the RS predicted by this model is either volatile or irreversible, since the filaments should quickly dissolve after removing the voltage, or never dissolve even with a reverse voltage (see the  Supporting Information (SI)).}

In this Letter, we propose a new \revis{nonvolatile and reversible} interfacial  RS mechanism, multiphase polarization (MP), fo LTO memristors and other similar systems made of  multiphase, ion-intercalation nanofilms enclosed by ion-blocking electrodes.

\textbf{Mechanism.} We begin by noting the following properties of LTO memristors derived from experimental data: (1) interfacial ET should  dominate the total resistance, since the effective conductivity of LTO4 and LTO7 measured by electrochemical impedance spectroscopy in Ref.\citenum{Gonzalez-Rosillo2020Lithium-BatterySeparation} is around $\SI{2e-11}{S/m}$, $\SI{6e-10}{S/m}$ at 30$\mathrm{^o C}$, which are small compared with the bulk values in the literature: $10^{-4}$-$10^{-11} \mathrm{S/m}$, $1$-$10^2 \mathrm{S/m}$ \cite{Young2013ElectronicAnode, Scharner1999EvidenceSpinel, Zhao2015APerspectives}; (2) the two electrodes should have different ET resistance due to different deposition temperatures; (3) LTO4 and LTO7 may not be pure (only Raman spectroscopy was used to estimate Li concentration) and both phases may exist in each memristor; (4) the ET rates usually strongly depend on local ion concentrations \cite{Sze2007PhysicsDevices, Bazant2013TheoryThermodynamics}.

\begin{figure}[t]
    \centering
    \includegraphics[width=\textwidth]{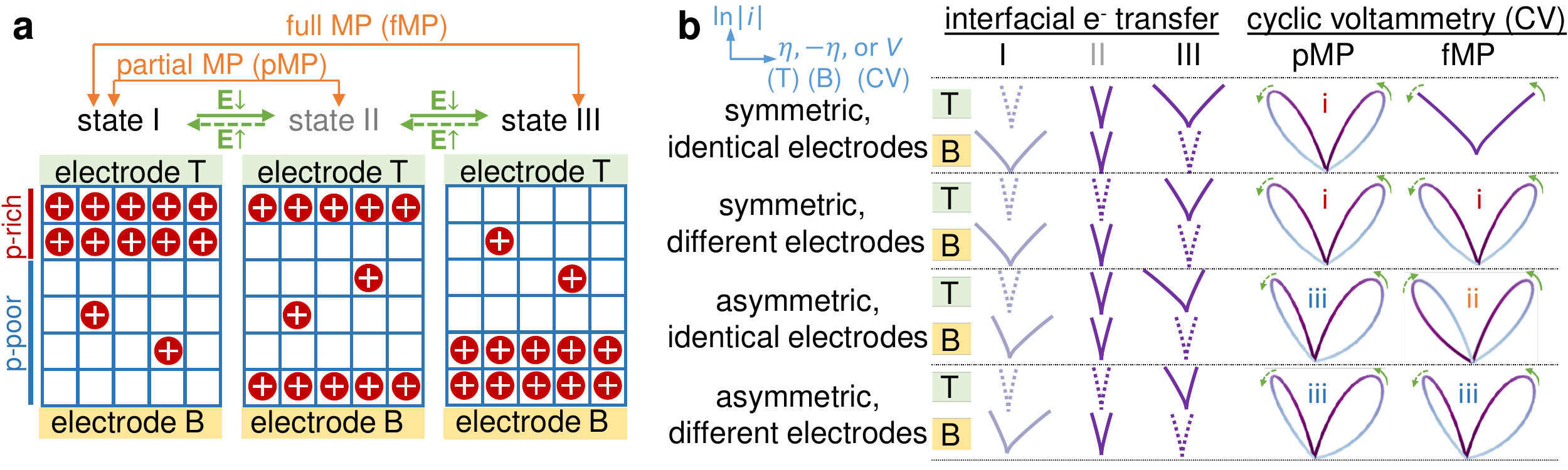}
    \caption{\revision{ (a) Schematics of the  phase re-distribution (state I $\leftrightarrow$ II $\leftrightarrow$ III)  by applied electric field ($\mathbf{E}$, whose direction is $\uparrow$ or $ \downarrow$) between ion-blocking electrodes, which process we refer to as multiphase polarization (MP). (b) The effect of MP on interfacial electron transfer (ET) rates and the resulting total resistance switching characterized by cyclic voltammetry, for different combinations of electrodes. We call an electrode symmetric if it does not conduct current primarily in one direction, and two electrodes  identical if they have the same ET rates given the same local concentration and overpotential. The two electrodes can be different due to materials and deposition conditions.  Each electrode contact is assumed to be more conductive and symmetric at higher concentration. Each curve represents the log of current ($\ln{|i|}$) v.s. interfacial overpotential $\eta$ or $-\eta$ for the top (T) or bottom (B) electrode at steady state, or v.s. the applied voltage $V$ for cyclic voltammetry. The ET rates shown by solid curves dominate total resistance. Three patterns of cyclic voltammetry are identified: (i) symmetric pattern with separated states, (ii) symmetric pattern with crossed states, (iii) asymmetric patterns.}      }
    \label{fig:schematic}
\end{figure}

These complex properties, which are not captured by existing theories, motivate us to propose the general MP mechanism, illustrated in  \autoref{fig:schematic}. 
We assume that the nanofilm conducts cations and electrons, while the electrodes only conduct electrons. Then cations tend to move along the electric field, but get blocked by the electrodes. Therefore, cations accumulate downstream the electric field, and get depleted on the other side. This phenomenon is called concentration polarization (CP), and is important in many electrochemical systems \cite{Mani2009OnAnalysis, Zangle2009OnStudy, Dydek2011OverlimitingMicrochannel, Dydek2013NonlinearModel, Nielsen2014ConcentrationMicrochannel, Mishchuk2010ConcentrationPhenomena, Andersen2012, Schlumpberger2015ScalableElectrodialysis, Mani2011DeionizationMicrostructures, Tian2020, Tian2020b, Tian2021ContinuousElectrodialysis}.
In multiphase materials, CP should first occur in each phase, and strong enough CP can change phase distribution, which we refer to as multiphase polarization (MP). \revision{ As shown in  \autoref{fig:schematic}(a),  a downward electric field may drive phase change at the bottom electrode first (I$\rightarrow $II), and then at the bottom electrode later (II$\rightarrow $III). We call the process I$\leftrightarrow$II partial MP (pMP), and I$\leftrightarrow$III full MP (fMP).} The phase distribution should be nonvolatile after removing the applied voltages \revision{since phases with different concentration can co-exist in multiphase materials}, unlike CP in homogeneous electrolytes. The phase distribution can further influence ET rates on electrodes and thus significantly influence the total resistance if it is dominated by contact resistance, as shown in \autoref{fig:schematic}(b). \revision{For example, if the two electrodes are symmetric (do not conduct electrons primarily in one direction) but the top electrode is more conductive than the bottom electrode given the same concentration and overpotential, then state I should have larger total resistance than state III, and thus fMP can lead to RS.  }  

To conclude, MP is an interfacial RS mechanism, limited by bulk ion diffusion, which shows multiple, non-volatile resistance states tunable by applied voltages in LTO memristors and other similar systems.


\textbf{Model.} Here, we develop an electrochemical phase-field model to quantitatively describe the mechanism, based on nonequilibrium thermodynamics of ion and electron transfer~\cite{ Bazant2013TheoryThermodynamics}. \revision{We neglect temperature variation which should be mild for the interfacial RS mechanism without hot filaments (see the SI), and we neglect  mechanical effects for ``zero-strain'' LTO \cite{Zhao2015APerspectives}. } Existing phase-field models of ion-intercalation materials in Li-ion batteries \cite{Bazant2013TheoryThermodynamics,Smith2017MultiphaseTheory,singh2008intercalation,tang2009model,Bai2011SuppressionDischarge,Cogswell2012CoherencyNanoparticles,Cogswell2013TheoryNanoparticles,Nadkarni2018InterplayNanoparticles,smith2017intercalation,Smith2017MultiphaseTheory,Nadkarni2019ModelingStorage,DeKlerk2017ExplainingModeling} and memristors \cite{Fraggedakis2020} consider only the dynamics of neutral Li$^+$-electron pairs. 
However, here we must account for large electric fields, and broken symmetry between ion and electron transfer at the electrode interfaces. 
As such, our model includes three charged species:  mobile, localized electrons (`n'), mobile, monovalent cations (`p'), and fixed, positively charged defects (`d'). The localized mobile electrons, also called ``small polarons" for their coupling with the polarized local environment (including nearby mobile ions), differ from band electrons by a much smaller but thermally-activated mobility and a non-thermally-activated number of mobile electrons. Such electrons are often found in mixed-valency transition metal oxides \cite{Cox2010TransitionProperties, Bosman1970Small-polaronOxides, Ellis2006SmallMobility, Zhou2006ConfigurationalLixFePO4, Mikkelsen1987ElectronReactions}, which include some of the most common ion-intercalation materials used in Li-ion batteries and other applications~\cite{sood2021electrochemical}. The fundamental constants we use include the Boltzmann constant $k_B$, temperature $T$, electron charge $e$, thermal voltage $V_T = k_B T/e$, and the Avogadro’s number $N_A$. We denote the concentration, valence, electrochemical potential, diffusivity, conductivity, flux density of mobile species $k$ ($=p,n$) as $c_k$, $z_k$ ($=\pm 1$), $\mu_k$, $D_k$, $\sigma_k$, $\mathbf{J}_k$, and the electric potential, current density, time, nanofilm thickness as $\phi$, $i$, $t$, $h$. We also define dimensionless variables $\tilde{c}_k = c_k/c_0$, $\tilde{\mu}_k = \mu_k/k_B T$, $\tilde{\phi} = \phi/V_T$, $\tilde{t} = t/\tau_D = t D_p^0/h^2$, $\tilde{J}_k = J_k h/D_k^0 c_0$, and $\tilde{i} = i/i_D = i/(D_n^0 c_0 N_A e/h)$, where $c_0$ and $D_k^0$ are constants. 

First,  we enforce electroneutrality:
\begin{equation}
    \tilde{c}_p = \tilde{c}_n - \tilde{c}_d = \tilde{c}.
    \label{eq:e_neutral}
\end{equation}
Then, we apply a regular solution model with Cahn-Hillard gradient expansion for the  two charge carriers ($k=p,n$) \cite{Zhou2006ConfigurationalLixFePO4, Bazant2013TheoryThermodynamics, Mebane2015ASolutions, Guyer2004PhaseEquilibrium, Guyer2004PhaseKinetics, Cahn1958FreeEnergy}:

\begin{equation}
    \tilde{\mu}_k = \ln \frac{\tilde{c}_k}{\tilde{c}_k^{max} - \tilde{c}_k} + \tilde{\mu}_k^0  + \Omega_k \tilde{c}^\rho + z_k \tilde{\phi} - \kappa_k \tilde{\nabla}^2 \tilde{c},
\end{equation}
where the five terms are: ideal entropy for a mixture of charge carriers and vacancies on $c_k^{max}$ sites,  constant standard energy,  mixing enthalpy, electrostatic potential energy,  and Cahn-Hillard gradient penalty  with $\kappa_n = \kappa_p = \kappa$. 
Next, we can use the homogeneous part of $\tilde{\mu}_p + \tilde{\mu}_n$ (with $\Omega=\Omega_n+ \Omega_p$ defined), 
to determine the thermal stability (see the SI), including the spinodal points $\tilde{c}_{s0}$, $\tilde{c}_{s1}$  and the binodal points $\tilde{c}_{b0}$, $\tilde{c}_{b1}$ (0: ion-poor phase; 1: ion-rich phase).  Compared with recent models of Li-ion battery materials \cite{DeKlerk2017ExplainingModeling, Fraggedakis2020, Nadkarni2019ModelingStorage, Smith2017MultiphaseTheory, Zhao2020LearningImages}, this analysis includes the  electrochemical potential of electrons, and uses an additional parameter `$\rho$' to adjust concentration-symmetry in the Gibbs free energy of mixing.

Next, we describe the transport of charge carriers by enforcing mass and charge conserva with the generalized Nernst-Planck equation for charge fluxes \cite{Maier2004PhysicalSolids, Bazant2013TheoryThermodynamics}
\begin{equation}
    r_k \frac{\partial \tilde{c}_k}{\partial \tilde{t}} + \tilde{\nabla} \cdot \tilde{\mathbf{J}}_k = 0, \ \ \ \mathbf{J}_k = - D_k c_k \nabla \tilde{\mu}_k,
    \label{eq:NP}
\end{equation}
where  $\tilde{\nabla} = h \nabla$, $r_p = 1$, $r_n= D_p^0/D_n^0$. The diffusivity with the excluded volume effects is $D_k = D_k^0  (1 - c_k/c_k^{max})$\revision{, and the corresponding ionic/electronic conductivity is $\sigma_k = D_k c_k e N_A/V_T$. We neglect field-dependence of $D_k^0$\cite{Menzel2019TheDevices} since we assume most of the electric potential falls at interfaces.}
The boundary conditions on the electrodes with outward normal vector $\mathbf{n}$ are: no penetration of ions $\mathbf{n} \cdot \tilde{\mathbf{J}}_p = 0$, conservation of electrons $\mathbf{n} \cdot \tilde{\mathbf{J}}_n = - \mathbf{n} \cdot \tilde{\mathbf{i}}$, and neutral wetting $\mathbf{n} \cdot \tilde{\nabla} \tilde{c} = 0$.

Finally, and critically, we must describe the ion-modulated ET rates at electrode interfaces.
\revision{At metal-semiconductor interfaces,  the ET rates can be described by diffusion, tunneling, or thermo-emission across a Schottky barrier, which can be significantly influenced by doping ions \cite{Tung2014TheHeight, Sze2007PhysicsDevices, Cowley1965SurfaceSystems}. However, unlike band conduction in traditional semiconductors, this work focuses on small polaron conduction in mixed-valency intercalation materials. In another words, the bulk ET in LTO and interfacial ET at LTO-Pt interface can be described by $\mathrm{Ti^{4+} + Ti^{3+} \rightleftharpoons Ti^{3+} + Ti^{4+}}$ and $\mathrm{Ti^{4+} + e^- (Pt) \rightleftharpoons Ti^{3+}}$, respectively. This ET mechanism, though in solids, is very analogous to the ET reactions in liquid solutions and at liquid-solid interfaces as described by Marcus theory \cite{Bai2014ChargeElectrodes, Mikkelsen1987ElectronReactions}, which involves a microscopic picture of solvent fluctuation and predicts curved Tafel plots (log current v.s. voltage) at large overpotentials as observed in experiments \cite{Marcus1964ChemicalTheory, Marcus1965OnReactions, Marcus1985ElectronBiology, Marcus1993ElectronExperiment, Chidsey1991FreeInterface, Henstridge2012Marcus-Hush-ChidseyReview}. Recently, Marcus theory was also applied to solid-state ET in carbon-coated lithium iron phosphate~\cite{Bai2014ChargeElectrodes}. However, the original Marcus theory and the asymmetric Marcus theory which has very limited applicability \cite{Laborda2012AsymmetricKinetics,Laborda2013AsymmetricVoltammetry, Zeng2015SimpleKinetics} cannot capture the significant asymmetry possibly seen at solid-solid interfaces. Then we notice the facts that  Marcus theory reduces to the phenomenological Butler-Volmer (BV) equation at small to medium overpotentials \cite{Fletcher2010TheTransfer, Fraggedakis2020TheoryKinetics, Bazant2013TheoryThermodynamics, Henstridge2012Marcus-Hush-ChidseyReview}, and BV equation with a series resistance can also lead to curved Tafel plots \cite{Smith2017MultiphaseTheory}. Though usually used for ion transfer, BV equation can be used for general Faraday reactions\cite{Bazant2013TheoryThermodynamics} including $\mathrm{Ti^{4+} + e^- (Pt) \rightleftharpoons Ti^{3+}}$. Actually a Schottky diode is also usually described by a BV-form formula in combination with a series resistance to fit the curved Tafel plots \cite{Sze2007PhysicsDevices, Lien1984AnResistance, Aubry1994SchottkyMethods}, though the ion-modulated exchange current should be different from the system we study here.}  As a first approximation to capture these diverse phenomena, we propose the following generalized BV equation,  which includes non-ideal thermodynamics to capture the concentration dependence and a series resistance $\tilde{R}_s$ to serve a similar role as Marcus theory to curve the Tafel plots \cite{Bazant2013TheoryThermodynamics, Smith2017MultiphaseTheory}: 
\begin{subequations}
\begin{equation}
    \mathbf{n} \cdot \tilde{\mathbf{i}} = \mathrm{Da} f(\tilde{c}; \alpha) g(\tilde{\eta} ; \alpha) e^{ -\alpha \kappa \tilde{\nabla}^2 \tilde{c} },
\end{equation}
\begin{equation}
    f(\tilde{c}; \alpha) =  (\tilde{c}_n/\tilde{c}_n^{max})^\alpha (1 - \tilde{c}_n/\tilde{c}_n^{max})^{1-\alpha} e^{\alpha \Omega_n \tilde{c}^\rho },
\end{equation}
\begin{equation}
    g(\tilde{\eta}; \alpha) = e^{(1-\alpha) \tilde{\eta}} - e^{-\alpha \tilde{\eta}},
\end{equation}
\label{eq:gBV}
\end{subequations}
where $\tilde{\eta} = \tilde{\mu}_n - \tilde{\mu}_e$ is the overpotential across the interface,
$\tilde{\mu}_e = \tilde{\mu}_e^0 - \tilde{\phi}_e$ is the Fermi energy in the electrode, and  $\tilde{\phi}_e^B =0$, $\tilde{\phi}_e^T = \tilde{V} -   \tilde{I} \tilde{R}_s  $ (where $\tilde{I}$ is the total current and in 1D $\tilde{I} = \tilde{i} = \tilde{\mathbf{i}}\cdot \mathbf{e}_x$, $\tilde{V}$ is the applied voltage, `T' and `B' represent the two electrodes at top and bottom).  We further assume $\alpha = \alpha_0 + (0.5-\alpha_0)\tilde{c}$ so that ET asymmetry disappears for large $\tilde{c}$ (Ohmic contact). Constants $\mathrm{Da}$ and $\alpha^0$ can be different for the two electrodes.   See the SI for derivation of \autoref{eq:gBV}.

\begin{figure*}[t!]
    \centering
    \includegraphics[width = \textwidth]{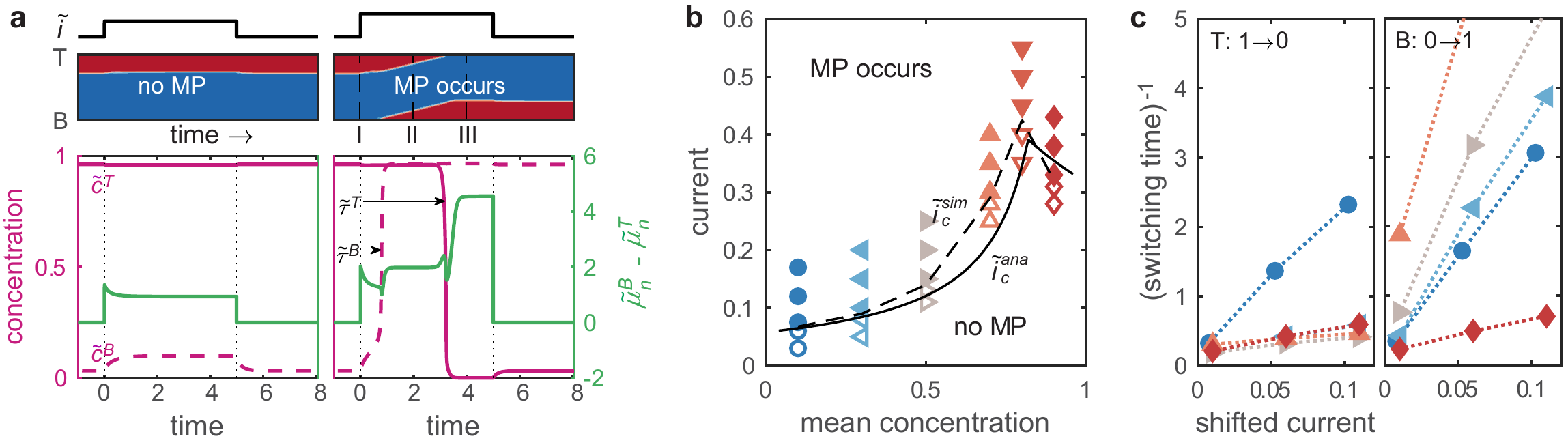}
    \caption{The switching current  and switching time  from simulations of step current response. (a) From top to bottom: applied current ($\tilde{i} = i/i_D$), phase distribution (contour map, red: ion-rich phase 1, blue: ion-poor phase 0), boundary concentrations ($\tilde{c}^T$, $\tilde{c}^B$), electrochemical potential drop ($\tilde{\mu}_n^B - \tilde{\mu}_n^T$),  along with time (\revis{$\tilde{t} = t/\tau_D$}) for two typical cases (with mean concentration $
    \tilde{c}_m = 0.3$, current plateau $\tilde{i}_{max} = 0.08, 0.12$). \revision{The schematics for the three typical states I, II, III during MP can be found in \autoref{fig:schematic}.} The switching time $\tilde{\tau}$ can be obtained from $\tilde{\tau}^T = \tilde{t}|_{\tilde{c}^T=0.5}$ and $\tilde{\tau}^B = \tilde{t}|_{\tilde{c}^B=0.5}$. (b) The critical current ($\tilde{i}_c$) for MP to occur indicated by simulations (filled and empty markers to indicate MP and no MP, dashed line for eye guidance) and theory (solid line), for different mean concentration ($\tilde{c}_m$). (c) The inverse of the switching time $1/\tilde{\tau}$  along with the shifted current ($\tilde{i} - \tilde{i}_c^{ana}$). The colors and markers to label concentrations in (b)(c) are consistent. }
    \label{fig:switching}
\end{figure*}
 
\begin{figure*}[t!]
     \centering
    \includegraphics[width =  \textwidth]{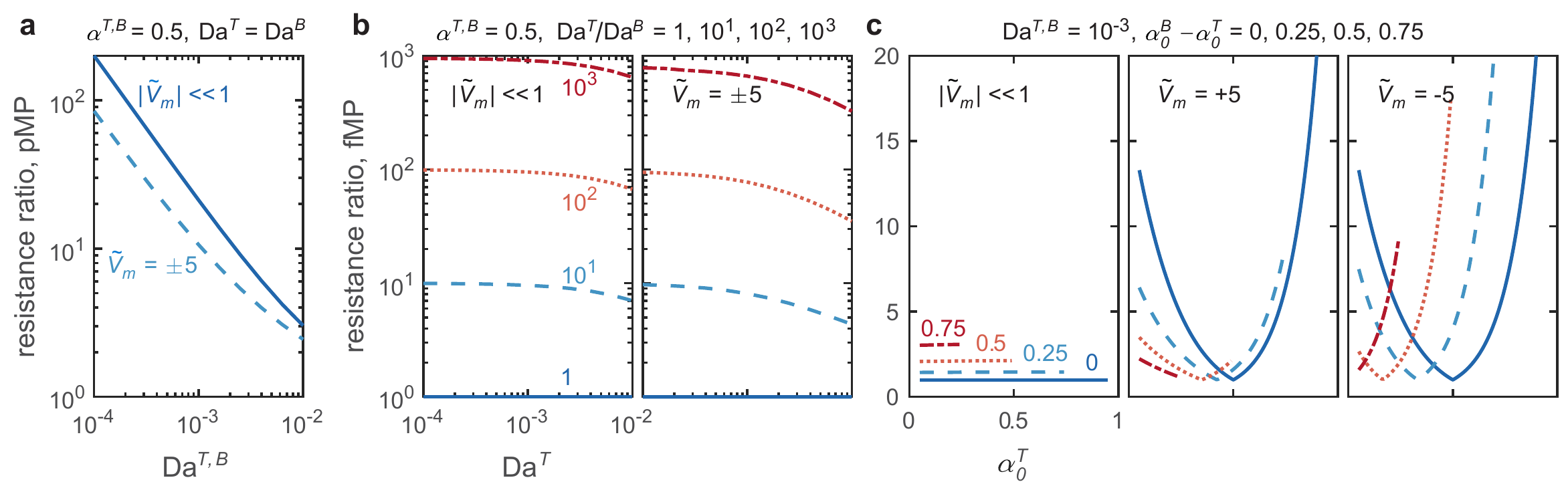}
    \caption{The resistance ratio ($\mathcal{R}$) measured at different voltages ($\tilde{V}_m = V_m/V_T$) due to (a) pMP and identical, symmetric electrodes, and (b) fMP  and symmetric electrodes with various rate constants Da, (c) fMP and asymmetric electrodes with the same Da. For pMP, $\mathcal{R} = \exp(|\ln \frac{\tilde{R}(\tilde{c}_{b1}, \tilde{c}_{b0})}{\tilde{R}(\tilde{c}_{b1}, \tilde{c}_{b1})} |)$. For fMP, $\mathcal{R} = \exp(|\ln \frac{\tilde{R}(\tilde{c}_{b1}, \tilde{c}_{b0})}{\tilde{R}(\tilde{c}_{b0}, \tilde{c}_{b1})} |)$. Here we choose $\tilde{c}_m=0.5$ to calculate the bulk resistance.  } 
    \label{fig:RS}
\end{figure*}

\begin{figure*}
    \centering
    \includegraphics[width = \textwidth]{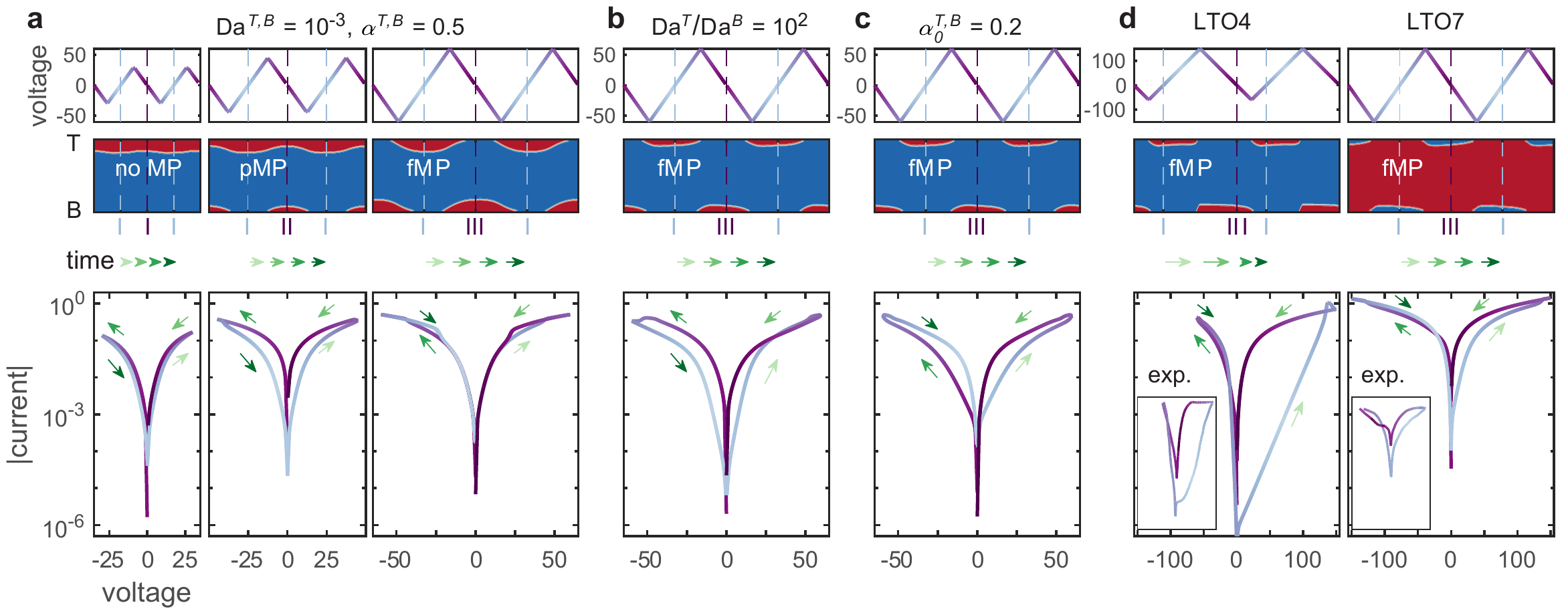}
    \caption{\revis{Simulation results of cyclic voltammetry for two cycles. From top to bottom: applied voltage ($\tilde{V} = V/V_T$) versus time ($\tilde{t} = t/\tau_D$), phase distribution (red: ion-rich phase 1, blue: ion-poor phase 0) versus time ($\tilde{t}$), and current $|\tilde{i}| = |i/i_D|$ versus voltage ($\tilde{V}$). The sweeping rate is 125 in dimensionless scale, or $\SI{50}{mV/s}$ if the diffusion time $\tau_D = \SI{64}{s}$. Four sets of electrodes are considered: (a) identical, symmetric electrodes with three maximum sweeping voltages, (b) symmetric, different electrodes, (c) identical, asymmetric electrodes, (d) different, asymmetric electrodes to fit for experiments of LTO4 and LTO7 memristors from Ref.\citenum{Gonzalez-Rosillo2020Lithium-BatterySeparation}. The dashed lines in the time evolution figures label three typical states, and schematics for state I, II, III can be found in \autoref{fig:schematic}.}  Interfacial parameters not shown in the figure: (b) $\mathrm{Da}^T =10^{-2}$, (c) $\mathrm{Da}^{T,B}=10^{-3}$, (d) LTO4:  $\mathrm{Da}^T = 10^{-3}$, $\mathrm{Da}^B = 10^{-6}$, $\alpha^T_0 = 0.3$, $\alpha^B_0 = 0.08 $, LTO7: $\mathrm{Da}^T = 10^{-1}$, $\mathrm{Da}^B = 10^{-3}$, $\alpha^T_0 = 0.05$, $\alpha^B_0 = 0.2 $. Mean concentration:  $\tilde{c}_m = 0.2, 0.1, 0.1, 0.1, 0.9$ for (a), (b), (c), (d)LTO4, (d)LTO7.  }
    \label{fig:cyclic}
\end{figure*}

\textbf{Dimensionless Results.}
We proceed to use our model to  analyze  RS performance of multiphase ion-intercalation nanofilms \cite{Waser2019IntroductionApplications} in terms of dimensionless variables. As an example to test the model, we choose parameters based on the LTO material \cite{ Zhao2015APerspectives, Young2013ElectronicAnode, Takami2011LithiumExtraction, Maier2004PhysicalSolids,  Scharner1999EvidenceSpinel,  Wagemaker2009Li-ion4+xTi5O12, Morgan2016VariationSurfaces} (see the SI):  $\tilde{c}_p^{max} = 1$, $\tilde{c}_n^{max}=5/3$, $\Omega= -12$,  $\rho = 0.7$,  $\kappa = \SI{1.1e-3}{}$, $\Omega_n = 20$, $D_p/D_n = 10^{-5}$, and $\tilde{c}_d = 0.01$, and $\tilde{R}_s = 100$.  Note that these values are all obtained/estimated from the experiments in the literature without fitting, and especially, $\tilde{R}_s$ is estimated from the reorganization energy of the small polarons.

We first analyze switching current and time, based on 1D simulations of MP in response to a step current (no need to consider ET here) as shown in \autoref{fig:switching}. \autoref{fig:switching}(a) shows two typical cases. Both cases reach steady state during the applied current, but only the larger current causes MP, in which case the time scale \revision{($\tilde{\tau}^T, \tilde{\tau}^B$)} for phase change at the two electrodes can be determined. \revision{The smaller one of $\tilde{\tau}^T, \tilde{\tau}^B$ is the time for pMP to occur, while the larger one is for fMP to occur. }Then the switching current and time for more cases with different averaged concentration $\tilde{c}_m$ and applied current $\tilde{i}$ are summarized in 
\autoref{fig:switching}(b)(c). The condition under which MP occurs can be explained as follows. In each phase, the concentration rises at the downstream of the electric field and falls on the other side. If this perturbation in either phase is large enough to cause phase change, MP occurs. We also derive an approximate, analytical expression for the critical current: 
\begin{equation}
    \tilde{i}_c^{ana} =  \min\left\{\frac{\mathcal{F}|_{\tilde{c}_{b0}}^{\tilde{c}_{s0} }}{\tilde{h}_0(\tilde{c}_m)},  \frac{\mathcal{F}|_{\tilde{c}_{s1}}^{\tilde{c}_{b1} }}{1 - \tilde{h}_0(\tilde{c}_m)}  \right\},  
\end{equation}
where $\mathcal{F}(\tilde{c})$ is the effective potential for current ($\tilde{i} = \tilde{\nabla}_x \mathcal{F}$), $\tilde{h}_0(\tilde{c})$  is an estimate of the occupation of the ion-poor phase 0  (see the SI). This expression is consistent with simulations, as shown in \autoref{fig:switching}(b). For the switching time, we find that its reciprocal is roughly proportional to the current, as shown in \autoref{fig:switching}(c). We also find that the end concentrations ($\tilde{c}=0.1, 0.9$) have similar switching time on the two sides, in which cases pMP is unlikely to be observed. And the $\tilde{c}=0.1$ case switches faster than $\tilde{c}=0.9$.  At medium concentrations, there is a big time window between pMP and fMP, which makes it possible to utilize pMP for RS.

Next, we analyze the resistance ratio (measured by voltage $\tilde{V}_m$) for different combinations of interfacial ET parameters.  Here we consider steady state and assume $|\tilde{V}_m|$ is not large enough to significantly perturb the bulk concentration profiles. If $|\tilde{V}_m| \ll 1$, the total resistance is $\tilde{R}(\tilde{c}^T, \tilde{c}^B) = \frac{1}{\mathrm{Da}^T f(\tilde{c}^T; \alpha^T)} + \frac{1}{\mathrm{Da}^B f(\tilde{c}^B; \alpha^B)} + \tilde{R}_b + \tilde{R}_s $, where $\tilde{R}_b $ is the bulk resistance and is usually much smaller than the other resistances (see the SI). If $|\tilde{V}_m| > 1$, we need to solve $\tilde{R}$ from the current balance. Then we can compare $\tilde{R}$ for different states and calculate the resistance ratio $\mathcal{R}$ ($\geq 1$ by definition), as shown in \autoref{fig:RS}. First, we need to know when MP can lead to RS ($\mathcal{R}>1$). Basically, for symmetric, identical electrodes, only pMP can lead to RS (see \autoref{fig:RS}(a) and $\mathrm{Da}^T/\mathrm{Da}^B=1$ case in \autoref{fig:RS}(b)). For other cases, fMP can also lead to RS. Then, to get a larger $\mathcal{R}$, the ET on the two electrodes should dominate the total resistance and be very different after MP. Therefore, larger $\Omega_n$ \revision{(more Fermi energy lift due to ion intercalation)} is preferred. In addition, for symmetric electrodes, $\mathcal{R}$ should decrease for larger Da, smaller Da ratios, larger $|\tilde{V}_m|$, as shown in \autoref{fig:RS}(a)(b). For asymmetric electrodes, the situation is more complex. Both the magnitude and sign of $\tilde{V}_m$ are very important, as shown in \autoref{fig:RS}(c). Only large enough $|\tilde{V}_m|$ can lead to $\mathcal{R}>1$ for identical, asymmetric electrodes and fMP (solid lines). Moreover, $\mathcal{R}$ for asymmetric, different electrodes depends on the sign of $\tilde{V}_m$.

Finally, we analyze the cyclic voltammetry behaviors.  As expected (\autoref{fig:schematic}), for symmetric and identical electrodes, only pMP can lead to RS, as shown in \autoref{fig:cyclic}(a). A Da ratio and fMP can lead to separated states (\autoref{fig:cyclic}(b)), while asymmetric $\alpha$ and fMP can lead to crossed states (\autoref{fig:cyclic}(c)),  around zero voltage.

\textbf{Comparison with experiments.}
For dimensional analysis, we use additional parameters: $c_0 = \SI{22}{M}$, $h = \SI{80}{nm}$, electrode area $S = \SI{500}{\mu m} \times \SI{500}{\mu m}$, $D_p^0 = \SI{1e-16}{m^2/s}$ (see the SI). The theory works well for the LTO memristors \cite{Gonzalez-Rosillo2020Lithium-BatterySeparation} in terms of the following three aspects. 

First, the theory predicts that the switching time is limited by the ion diffusion time ($\tau_D = \SI{64}{s}$), though it can be reduced by over ten times by increasing the current (\autoref{fig:switching}). Therefore, we predict switching time in seconds, which is consistent with experiments. 

Second, without any fitting parameters, our theory indicates that LTO4 memrsitor should have faster switching than LTO7 and need less current (case $c=0.1, 0.9$ in \autoref{fig:switching}), which are also consistent with  Figure 2(c)(d) in Ref.\citenum{Gonzalez-Rosillo2020Lithium-BatterySeparation}.  

Finally, by choosing proper interfacial parameters, we can obtain cyclic voltammetry patterns similar to experiments (Figure 2(a)(b) in Ref.\citenum{Gonzalez-Rosillo2020Lithium-BatterySeparation}), as shown in \autoref{fig:cyclic}(d). \revis{Note that the dimensionless current at low and high voltage is mainly determined by Da and $\tilde{R}_s$, respectively; and the dimensional current scale is mainly determined by the diffusion current $i_D$.  }  
Since electronic conductivity can vary by magnitudes due to defects \cite{Young2013ElectronicAnode, Scharner1999EvidenceSpinel}, we may assume LTO7 conductivity to be $\sim \SI{0.005}{S/m}$ to quantitatively fit the experimental current in \autoref{fig:cyclic}(d) (see the SI).

\revision{However, the 1D simple picture cannot predict the numerous resistance states and finite retention time observed in experiments. These issues may be explained by 2D or 3D phase nucleation, which may also reduce the switching current significantly..}

\textbf{Discussion and perspectives.}
Our model provides some fresh insights into the optimization and possible alternative designs for MP-based memristors.  

\begin{figure}
    \centering
    \includegraphics[width = 0.5\textwidth]{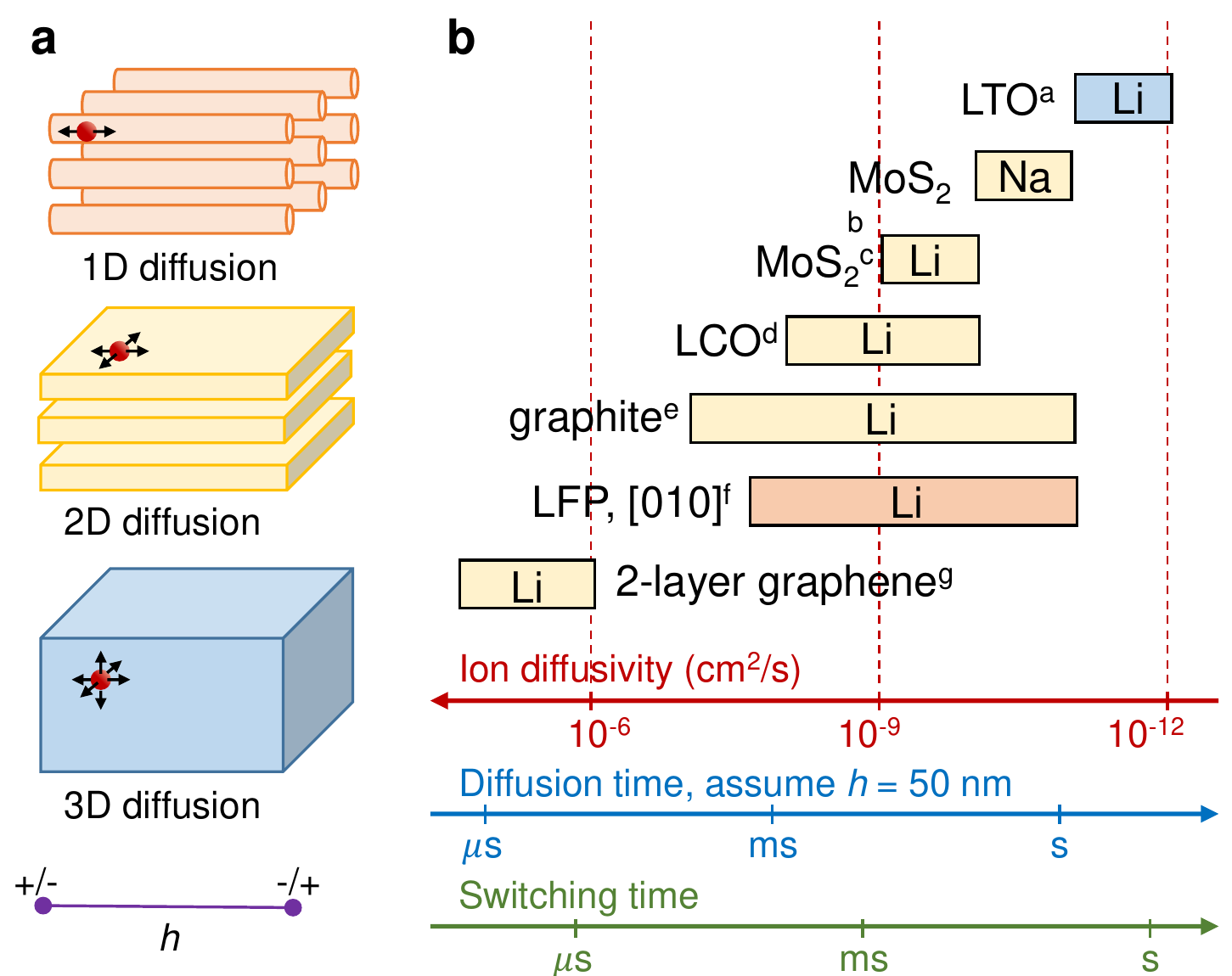}
    \caption{The scales of ion diffusivity (at room temperature), diffusion time, and switching time by multiphase polarization, for several ion-intercalation materials. We assume the length scale between electrodes ($h$) to be 50 nm to calculate the diffusion time, and we estimate the switching time to be one-tenth of the diffusion time at certain current. The colors labeling diffusion path dimensions in (a) are consistent with the colors for different materials in (b). References: $^a$Ref.\citenum{Nitta2015Li-ionFuture}, $^b$Ref.\citenum{Li2015EnhancingNanocomposites},  $^c$Refs.\citenum{Hu2016MoS2Batteries, Santa-Ana1995TemperatureMoS2},  $^d$Ref.\citenum{Park2010ABatteries},  $^e$Ref.\citenum{Nitta2015Li-ionFuture}, $^f$Ref.\citenum{Malik2010ParticleDiffusivity}, $^g$Ref.\citenum{Kuhne2017UltrafastGraphene}. }
    \label{fig:perspectives}
\end{figure}

First,  the switching time of existing LTO memristors (on the order of seconds) is too long for the requirements of in-memory computing, including neuromorphic artificial synapses ($\leq \SI{}{\mu s}$-$\SI{}{ms}$)\cite{Prezioso2015TrainingMemristors} and digital computing ($\leq \SI{}{ns}$) \cite{Ielmini2018In-memoryDevices}. The performance can be improved by changing from LTO to intercalation materials with high ion mobility, decreasing nanofilm thickness, or increasing current. For example, we can choose intercalation materials with lower-dimensional diffusion paths (\autoref{fig:perspectives}(a)), e.g., nano-sized LFP (defects-sensitive 1D paths) \cite{Malik2010ParticleDiffusivity}, layered materials (2D paths) \cite{Zhou2021LayeredMaterials} like MoS$_2$ \cite{Li2015EnhancingNanocomposites, Hu2016MoS2Batteries,Santa-Ana1995TemperatureMoS2},  LCO \cite{Park2010ABatteries}, graphite \cite{Nitta2015Li-ionFuture, Li2019IntercalationBeyond}, and even 2D materials (with only a few layers)\cite{Stark2019Intercalation2D, Kuhne2017UltrafastGraphene}. The ion diffusivity for these examples are shown in \autoref{fig:perspectives}(b). Note that these intercalation materials all have phase separation and strong concentration dependence of conductance (thus usually of contact resistance), which is necessary for MP-induced RS. We then assume the nanofilm thickness is 50 nm and the switching time is one-tenth of the diffusion time, and put the axes for diffusion and switching time in \autoref{fig:perspectives}(b). This is a conservative estimation, but note that we need a thick enough nanofilm to allow co-existence of different phases, and a too large current may lead to problems like Li plating. As we can see, we should be able to obtain switching time of $\SI{}{\mu s}$-$\SI{}{ms}$ which is sufficient for neuromorphic computing.

In addition, the theory implicitly indicates infinite retention time, without state decay by diffusion which usually occurs for traditional interfacial mechanisms  \cite{Sassine2016Devices}. The experimental finite retention time may come from rich-phase detachment from electrodes  due to material heterogeneity, surface wetting, or thermal fluctuation. This can be (partially) avoided by surface processing and scale-down, which should also help increase recyclability and reduce stochasticity. Finally, scale-down should be the primary way to  reduce power consumption. 

More discussion can be found in the SI.

\textbf{Conclusion.}
In this work, we have proposed and modeled a new interfacial resistive-switching mechanism, multiphase polarization, for a system composed of a multiphase, ion-intercalation nanofilm sandwiched by two ion-blocking electrodes.  This model is the first to qualitatively explain the complex RS dynamics of LTO memristors, and it provides insights for device optimization and new designs. Future theoretical work could account for 2D or 3D phase nucleation at interfaces \cite{Granasy2007PhaseNucleation,Cogswell2013TheoryNanoparticles},  thermal and mechanical effects~\cite{Cogswell2012CoherencyNanoparticles,Nadkarni2018InterplayNanoparticles}, and multi-stage phase separation~\cite{smith2017intercalation}.

\begin{acknowledgement}
This work was supported by a grant from Ericsson. 
The authors would like to thank Danniel Cogswell, Willis O'Leary,  Dimitrios Fraggedakis, Moran Balaish, Drew Buzzell,   Jennifer Rupp,  Ju Li, and Pedro de Souza for helpful discussions. 
\end{acknowledgement}

\begin{suppinfo}
The Supporting information is available free of charge at \url{https://pubs.acs.org/doi/10.1021/acs.nanolett.2c01765}.

Details on the dielectric breakdown model; temperature effect; thermal stability analysis;  derivation of the generalized Butler-Volmer equation, threshold current,  and resistance ratio; parameters; numerical method; additional discussion.  
\end{suppinfo}

\bibliography{ref_memristor.bib, ref_electrochem.bib, ref_shockED.bib, ByHand.bib}

\section{Graphic Abstract}
\begin{figure}
    \centering
    \includegraphics[width=0.5 \textwidth]{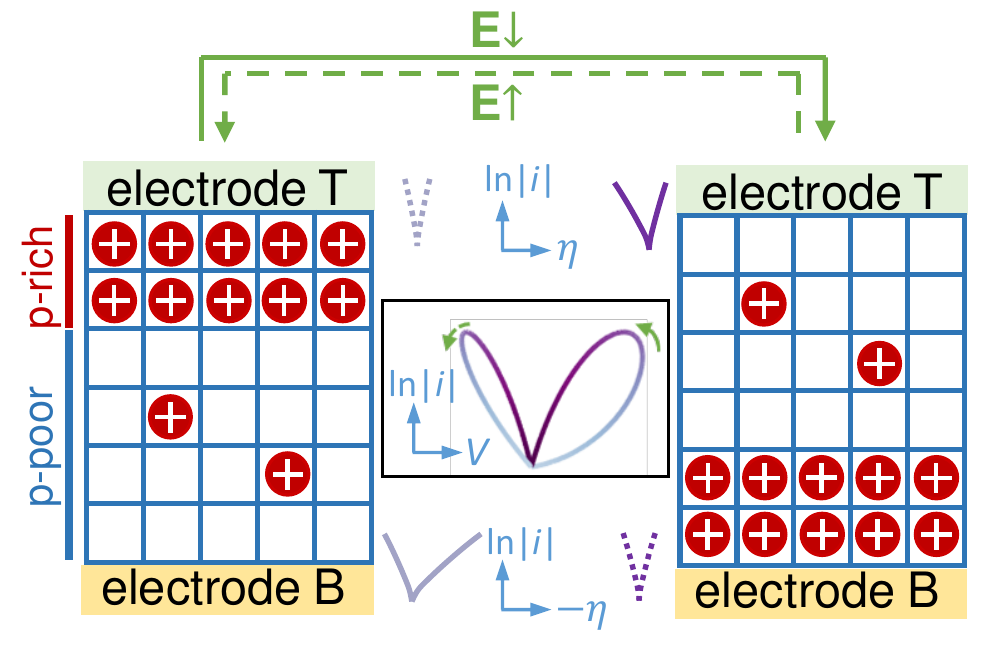}
    \label{fig:my_label}
\end{figure}
\end{document}